\documentstyle[12pt]{article}

\advance\textheight by \topskip
\begin{document}
\tolerance=100000
\input feynman
\thispagestyle{empty}
\setcounter{page}{0}

\newcommand{\be}{\begin{equation}}
\newcommand{\ee}{\end{equation}}
\newcommand{\br}{\begin{eqnarray}}
\newcommand{\er}{\end{eqnarray}}
\newcommand{\ba}{\begin{array}}
\newcommand{\ea}{\end{array}}
\newcommand{\bi}{\begin{itemize}}
\newcommand{\ei}{\end{itemize}}
\newcommand{\bn}{\begin{enumerate}}
\newcommand{\en}{\end{enumerate}}
\newcommand{\bc}{\begin{center}}
\newcommand{\ec}{\end{center}}
\newcommand{\ul}{\underline}
\newcommand{\ol}{\overline}
\newcommand{\eebbww}{$e^+e^-\rightarrow b\bar b W^+W^-$}
\newcommand{\eebbzz}{$e^+e^-\rightarrow b\bar b Z^0Z^0$}
\newcommand{\eebbvv}{$e^+e^-\rightarrow b\bar b VV$}
\newcommand{\bb}{$ b\bar b$}
\newcommand{\ttb}{$ t\bar t$}
\newcommand{\ar}{\rightarrow}
\newcommand{\sm}{${\cal {SM}}$}
\newcommand{\Dir}{\kern -6.4pt\Big{/}}
\newcommand{\Dirin}{\kern -10.4pt\Big{/}\kern 4.4pt}
\newcommand{\DDir}{\kern -7.6pt\Big{/}}
\newcommand{\DGir}{\kern -6.0pt\Big{/}}

\def\Ord{\buildrel{\scriptscriptstyle <}\over{\scriptscriptstyle\sim}}
\def\OOrd{\buildrel{\scriptscriptstyle >}\over{\scriptscriptstyle\sim}}
\def\pl #1 #2 #3 {{\it Phys.~Lett.} {\bf#1} (#2) #3}
\def\np #1 #2 #3 {{\it Nucl.~Phys.} {\bf#1} (#2) #3}
\def\zp #1 #2 #3 {{\it Z.~Phys.} {\bf#1} (#2) #3}
\def\pr #1 #2 #3 {{\it Phys.~Rev.} {\bf#1} (#2) #3}
\def\prep #1 #2 #3 {{\it Phys.~Rep.} {\bf#1} (#2) #3}
\def\prl #1 #2 #3 {{\it Phys.~Rev.~Lett.} {\bf#1} (#2) #3}
\def\mpl #1 #2 #3 {{\it Mod.~Phys.~Lett.} {\bf#1} (#2) #3}
\def\rmp #1 #2 #3 {{\it Rev. Mod. Phys.} {\bf#1} (#2) #3}
\def\xx #1 #2 #3 {{\bf#1}, (#2) #3}
\def\preprint{{\it preprint}}

\begin{flushright}
{\large DFTT 69/94}\\
{\large DTP/95/02}\\
{\rm December 1994\hspace*{.5 truecm}}\\
\end{flushright}

\vspace*{\fill}

\begin{center}
{\Large \bf Heavy
Higgs production and decay via $e^+e^-\rightarrow Z^0 H^0
\rightarrow b\bar bZ^0Z^0$
and irreducible backgrounds at Next Linear
Colliders.\footnote{Work supported in part by Ministero
dell' Universit\`a e della Ricerca Scientifica.\\[4. mm]
E-mails: Moretti@to.infn.it;
Stefano.Moretti@durham.ac.uk.}}\\[2.cm]
{\large Stefano Moretti}\\[0.5 cm]
{\it Dipartimento di Fisica Teorica, Universit\`a di Torino,}\\
{\it and I.N.F.N., Sezione di Torino,}\\
{\it Via Pietro Giuria 1, 10125 Torino, Italy.}\\[0.5cm]
{\it Department of Physics, University of Durham,}\\
{\it South Road, Durham DH1 3LE, United Kingdom.}\\[0.75cm]
\end{center}

\vspace*{\fill}

\begin{abstract}
{\normalsize
\noindent
The complete matrix element for $e^+e^-\ar b\bar bZ^0Z^0$ has been
computed at tree--level and applied to
$Z^0H^0$--production followed by $Z^0\ar b\bar b$ and $H^0\ar Z^0Z^0$,
including all the irreducible
background, at Next Linear Colliders. We find that,
assuming flavour identification of the $Z^0$--decay products,
this channel, together with
$e^+e^-\ar b\bar bW^+W^-$ in which $Z^0H^0\ar (b\bar b)(W^+W^-$), can be
important for the study of the parameters of the Standard
Model Higgs boson over the heavy mass range $2M_{Z^0}\Ord M_{H^0}\Ord
2{m_t}$.}
\end{abstract}

\vspace*{\fill}
\newpage

\subsection*{Introduction}

Despite the innumerable phenomenological successes of the Standard Model (\sm),
an essential ingredient is still missing: the discovery of the
Higgs boson $H^0$. This particle plays a crucial role in generating
the spontaneous symmetry breaking of the $SU(2)_L\times U(1)_Y$
gauge group of the electroweak interactions, and in ensuring
the renormalizability of the whole theory.
We know that the $H^0$ is supposed to
be a ${\cal {CP}}$--even neutral scalar boson, we know its couplings to the
other elementary particles, but
no prediction on its mass (i.e., $M_{H^0}$) can theoretically be done.\par
However, an upper bound of approximately 1 TeV
(from perturbative unitarity arguments \cite{unitarity}) is expected, whereas
a lower limit can be derived from
current experiments at LEP I. In fact,
from unsuccessful searches for $e^+e^-\ar
Z^0\ar Z^{0*}H^0$ events at the $Z^0$--peak,
one can deduce the bound $M_{H^0}\OOrd 60~{\mathrm {GeV}}$ \cite{limSM}.\par
Assuming the above mass range, various studies on the feasibility of its
detection by the next generation of high energy
machines have been carried out, both at hadron colliders
\cite{guide,LHC,SSC,Tevatron} and
at the $e^+e^-$ ones \cite{guide,LepII,NLC,ee500,LC92,JLC}.\par
On the basis of the expected center--of--mass ({\tt c.m.}) energies,
luminosities, detector performances of these accelerators
and of the predicted cross sections and branching ratios,
it has been definitively demonstrated that, if the $H^0$ is in the
mass region $M_{H^0} \Ord M_{Z^0}$ (i.e., light Higgs), it can be
discovered
at LEP II  (with $\sqrt s_{ee}=160\div200$ GeV) in a large variety of
channels \cite{LepII}.
For a larger mass Higgs,
a $pp$ colliders like the LHC ($\sqrt s_{pp}=14$ TeV)
and/or an $e^+e^-$ accelerator like the Next Linear Collider (NLC, with $\sqrt
s_{ee}=300\div 1000$ GeV)
is needed. Even though
at the LHC the mass range $M_{Z^0}\Ord M_{H^0}
\Ord 130$ GeV is quite difficult to cover since
in this case the Higgs boson mainly
decays to $b\bar b$--pairs (signature which has a huge QCD background
if $b$--quarks cannot be recognized), nevertheless, it should be possible to
detect it
in the rare $\gamma\gamma$--decay mode \cite{gamgam}
via the associated production with a $W^\pm$ boson \cite{gny,wh}
or a $t\bar t$--pair \cite{rwnz,tth}. At the LHC,
for $M_{H^0} \OOrd 130$ GeV, the
``gold-plated'' four--lepton mode (i.e.,
$H^0\rightarrow Z^0Z^0\rightarrow \ell^+\ell^-\ell^+\ell^-$),
via various production channels,
remains the clearest signature \cite{LHC,SSC}.
At NLCs, with $\sqrt s_{ee}=300\div 500$ GeV,
the Higgs detection is possible
over the whole intermediate mass range (i.e.,
$M_{Z^0}\Ord M_{H^0}\Ord 2 M_{W^\pm}$) \cite{BCDKZ}, via
the bremsstrahlung reaction  $e^+e^-
\rightarrow Z^{0*}\rightarrow Z^0H^0$ \cite{bremSM} and/or the fusion processes
$e^+e^-\rightarrow \bar\nu_e\nu_eW^{\pm*}
W^{\mp*}(e^+e^-Z^{0*}Z^{0*})\rightarrow\bar\nu_e\nu_e
(e^+e^-)H^0$ \cite{fusionSM}. If
$\sqrt s_{ee}\OOrd500$ GeV, a heavy Higgs (i.e., $M_{H^0}\OOrd 2M_{W^\pm}$,
and mainly produced via the fusion processes),
can be detected
via the four--$jet$ modes
$H^0\rightarrow W^\pm W^\mp,Z^0Z^0\rightarrow jjjj$
as well as via the $4\ell$--decay
\cite{BCKP,4jet}. Finally, signatures that can be disentangled
through $b$--tagging \cite{SDC}, must also be added
to the mentioned channels:
such as, e.g., at the LHC, $t\bar tH^0$ production, with one $t(\bar t)$
decaying semileptonically and $H^0\rightarrow b\bar b$, with 80 GeV $\Ord
M_{H^0} \Ord$ 130 GeV \cite{btagg}.\par
In a recent study \cite{eezh}, we presented an
analysis of the Bjorken reaction $e^+e^-\ar Z^0H^0$ in the case
of a heavy Higgs decaying to $W^+W^-$--pairs and with $Z^0\ar b\bar b$,
and of all the $b\bar b W^+W^-$ irreducible background,
assuming flavour identification of the $Z^0$--decay products.
We emphasized in that work the importance of $b$--tagging
the weak boson $Z^0$, as this could be one of the most efficient ways
of detecting it, since this channel is free from $W^\pm$--decay
backgrounds, has a branching ratio approximately five times larger
than that one of $Z^0$ decaying to $\ell^+\ell^-$--pairs (with $\ell=e$, $\mu$
or $\tau$),
and is
comparable to the fraction of invisible decays $Z^0\rightarrow \nu\bar\nu$.
This, obviously, relies on the expected efficiencies and purities
for $b$--tagging at NLCs \cite{tag}.\par
In ref.~\cite{eezh} we found that, after carrying a missing mass analysis
\cite{GHS} on \eebbww, there are values of the Higgs mass
for which a simple cut on the invariant
mass $M_{b\bar b}$ is sufficient in order to
completely eliminate the irreducible background (which is dominated by
\ttb--production and decay) when
the double distributions
$d\sigma/dM_{b\bar b}/dM_{W^+W^-}$ of signal and background events do not
overlap in the plane
$(M_{b\bar b},M_{W^+W^-})$. Otherwise, further cuts
based on the kinematics of the \ttb--background
are needed, and these still maintain
an acceptable number of events from Higgs production.\par
The missing mass method has the useful feature of being
independent of assumptions on the $H^0$--decay modes, but this
means that as the $b\bar b W^+W^-$ events enter
in the missing mass distribution so should the
$b\bar b Z^0 Z^0$ ones,
and with a quite large component of signal
$H^0\rightarrow Z^0Z^0$ if compared to $H^0\rightarrow W^+W^-$,
since the $Z^0Z^0$--branching ratio
is only a factor of two/three less than
the $W^+W^-$--one in the heavy Higgs mass region
(with $M_{H^0}\Ord 2m_t$). Because of this  ``inclusive''
analysis on the decay products of the Higgs boson,
both signals and irreducible backgrounds of both the above processes must be
then considered at the same time\footnote{Moreover, in the
range $2M_{W^\pm}\Ord M_{H^0}\Ord2{m_t}$,
$H^0\ar W^+W^-$ and $H^0\ar Z^0Z^0$ are the only
relevant branching fractions.}.
Therefore, we retain that a complete
study, which includes $e^+e^-\ar b\bar b Z^0Z^0$
as well as $e^+e^-\ar b\bar b W^+W^-$,
is needed in order to definitively establish the feasibility of all the
foreseen
measurements of the Higgs boson parameters, if
the missing mass analysis is adopted and the $Z^0$ is assumed to decay
to $b\bar b$--pairs, with these tagged by vertex detectors.\par
As already done in ref.~\cite{eezh} we do not include in our
computations the beam energy spread resulting from
bremsstrahlung and beamsstrahlung effects, therefore,
as explained there,
one has to expect both the number of events and their statistical significance
to be slightly higher than those ones we predict here.\par
In this letter, using the full matrix element for the process
$e^+e^-\ar b\bar bZ^0Z^0$ we study the production of a heavy \sm\ Higgs (i.e.,
$M_{H^0}\ge 2M_{W^\pm}$)
via the Bjorken bremsstrahlung reaction $e^+e^-\ar Z^0H^0$, followed by the
decays $Z^0\ar b\bar b$ and $H^0\ar Z^0Z^0$, and of all the irreducible
$b\bar bZ^0Z^0$ background. Moreover, we present final results in which
both the rates for
$b\bar b W^+W^-$ and $b\bar b Z^0Z^0$ are added together.\par
Following the track of ref.~\cite{eezh}, we give details of the
calculation in section II, while in section
III we present and discuss the results.
Finally, section IV is devoted to our conclusions.

\subsection*{Calculation}

All the Feynman diagrams describing the process \eebbzz\ at tree--level
are shown in fig.~1, where graphs in which $Z^0$'s can be
exchanged (i.e., when they do not come from the same vertex)
must be counted twice (exchanging the corresponding quadrimomenta).
The matrix element has been computed using the method of ref. \cite{hz} and
we have checked the {\tt FORTRAN} code for BRS
invariance \cite{BRS} and compared it with a second one, produced by
MadGraph \cite{tim} and using the package HELAS \cite{helas}\footnote{Running
then only the first one for producing results.}.\par
The following numerical values of the parameters were adopted:
$M_{Z^0}=91.1$ GeV, $\Gamma_{Z^0}=2.5$ GeV,
$\sin^2 (\theta_W)=0.23$,
$m_b=5.0$ GeV and $\alpha_{em}= 1/128$. For
the Higgs width (i.e., $\Gamma_{H^0}$) we have adopted the tree--level
expression, and
we have not included effects of the width of the final state $Z^0$'s.\par
A few thoughts will now be devoted
to the procedure adopted for the integration
of the matrix element over the phase space.
In order to control the interplay between the various peaks
which appear in the integration domain when all tree--level
contributions are kept into account,
we have split the Feynman amplitude squared into
a sum of different (non gauge invariant) terms, each of which
corresponds to
the modulus squared of the resonant diagrams
(for each possible resonance) and,
eventually, their interference with other channels
\cite{eezh}.
In a similar way, the contribution of non--resonant diagrams
must also be considered.\par
Explicitly, in the case of the process \eebbzz\, with $M_{H^0}>2M_{Z^0}$,
we have $H^0\ar b\bar bZ^0$, $Z^0\ar b\bar b$,
$H^0\ar Z^0Z^0$, $Z^0H^0\ar (b\bar b)(Z^0Z^0)$ and $H^0\ar b\bar b$ resonances,
via the five channels (see fig.~1)\footnote{Diagrams
with exchanged $Z^0$'s are here implied.}:
\vskip0.5cm
\noindent
${M}_1:\quad\quad\quad  H^0\ar b\bar bZ^0\quad\quad{\mathrm
{diagrams}~\#~11,12,18},$

\noindent
${M}_2:\quad\quad\quad  Z^0\ar b\bar b\quad\quad{\mathrm
{diagrams}~\#~4,5,6~(with}~Z^0-{\mathrm {propagators)}},$

\noindent
${M}_3:\quad\quad\quad  H^0\ar Z^0Z^0\quad\quad{\mathrm {diagrams}~\#~15,16},$

\noindent
${M}_4:\quad\quad\quad  Z^0H^0\ar (b\bar b)(Z^0Z^0)\quad\quad{\mathrm
{diagram}~\#~17},$

\noindent
${M}_5:\quad\quad\quad  H^0\ar b\bar b\quad\quad{\mathrm {diagrams}~\#~13,14}.$
\vskip0.5cm
\noindent
Diagrams \# 1--3, 7--10, and 4--6 (with $\gamma$~--~propagators)
constitute the sixth (non--resonant) channel (${M}_6$).
Obviously, if $M_i$ indicates the sum of the diagrams entering
in the $i$--th channel, one has
\be\label{sum} M_{tot}=\sum_{i=1}^{6} M_i,\ee
where $M_{tot}$ is the total Feynman amplitude.
In squaring equation (\ref{sum}) we take the
combinations\footnote{This in order to minimize the errors coming from the
multi--dimensional integrations over the phase space, when we need to integrate
interferences between channels with and without (or different) resonances.}
\be\label{m1} {\cal M}_{1}^2=|M_1|^2,\quad\quad{\cal
M}_{2}^2=|M_2|^2+2\Re[M_2M_4^*], \ee
\be\label{m2} {\cal M}_{3}^2=|M_3|^2+2\Re[M_3M_4^*], \ee
\be\label{m3} {\cal M}_{4}^2=|M_4|^2,\quad\quad{\cal M}_{5}^2=|M_5|^2, \ee
\br\label{m4} {\cal M}_{6}^2&=&|M_6|^2\nonumber \\
	          & &   +2\Re[M_1M_2^*]+2\Re[M_1M_3^*]
                        +2\Re[M_1M_4^*]+2\Re[M_1M_5^*]\nonumber \\
                  & &   +2\Re[M_1M_6^*]+2\Re[M_2M_3^*]
                        +2\Re[M_2M_5^*]+2\Re[M_2M_6^*]\nonumber \\
                  & &   +2\Re[M_3M_5^*]+2\Re[M_3M_6^*]
                        +2\Re[M_4M_5^*]+2\Re[M_4M_6^*]\nonumber \\
                  & &   +2\Re[M_5M_6^*], \er
where $\Re(x)$ represents the real part of $x$, and with
\be\label{sum2} |{M}_{tot}|^2=\sum_{i=1}^{6} {\cal M}_i^2,\ee
where $|{M}_{tot}|^2$ is the total Feynman amplitude squared.\par
Then, to obtain an integrand function smoothly dependent
on the integration variables, for each contribution
in the matrix element
(\ref{sum2}) containing a resonance we make the change
\begin{equation}
p^2-M^2=M\Gamma\tan\theta,
\end{equation}
this factorizes the Jacobian
\begin{equation}
{\mathrm d}p^2=
\frac{1}{M\Gamma}[(p^2-M^2)^2+M^2\Gamma^2]{\mathrm d}\theta,
\end{equation}
which removes the dependence on the Breit--Wigner peaks appearing
in the ${\cal M}_i^2$ terms.
Here, $p$, $M$ and $\Gamma$
stand for the quadrimomentum, the mass and the width of the resonance,
respectively.
Then, we separately integrated
the various contributions (\ref{m1})--(\ref{m4})
by VEGAS \cite{vegas}, using
an appropriate phase space for each.\par
Finally, throughout this paper we
adopt an integrated luminosity ${\cal L}=10$
pb$^{-1}$ and we assume that only one $b$--$jet$ is tagged, with
an efficiency $\epsilon_b=1/3$ (i.e.,
$\epsilon_b$ is the probability for a $b$--quark
to satisfy a given set of tagging requirements). Therefore,
the probability of
tagging at least one $b(\bar b)$ out of a $b\bar b$--pair is
$P_{1}=1-(1-\epsilon_{b})^2=5/9\approx0.56$. In principle, we should
consider here the fact that there are also the other two $Z^0$'s in the
event, one or both of which can decay to $b\bar b$--pairs.
To this aim, we express
$P_n=1-(1-\epsilon_{b})^{2n}$ to be
the probability of tagging at least one
$b(\bar b)$ out of $n$ $b\bar b$--pairs, and we ``roughly'' split the
total cross section $\sigma(e^+e^-\ar b\bar b Z^0Z^0)$ into three
contributions:
$\sigma_3=\sigma(e^+e^-\ar b\bar b Z^0Z^0)\times[BR(Z^0\ar b\bar b)]^2
\times\left(\frac{\delta_{b\bar b,b\bar b,b\bar b}}{\delta_{Z^0Z^0}}\right)$,
$\sigma_2=\sigma(e^+e^-\ar b\bar b Z^0Z^0)\times[2BR(Z^0\ar b\bar b)]
\times\left(\frac{\delta_{b\bar b,b\bar b}}{\delta_{Z^0Z^0}}\right)$ and
$\sigma_1=\sigma(e^+e^-\ar b\bar b Z^0Z^0)-\sigma_2-\sigma_3$, corresponding to
the case of three, two and
one final $b\bar b$--pairs from $Z^0$--decays, respectively.
Here, $BR(Z^0\ar b\bar b)\approx0.15$ is the $Z^0$--branching ratio
into $b$--quarks, whereas $\delta_{Z^0Z^0}(\delta_{b\bar b,b\bar b})
[\delta_{b\bar b,b\bar b,b\bar b}]=1/2(1/4)[1/36]$ is the
$1/k!$ factor for each $k$--uple of identical particles
(since we integrated over the whole phase space)
in $b\bar b X Y(b\bar b b\bar b X)[b\bar b b\bar b b\bar b]$
final states, with $X$ and $Y$
not representing $b$--particles. Then, we expect
the efficiency of tagging at least one $b(\bar b)$ out of
all the possible final signatures of $b\bar bZ^0Z^0$ events to be
$P_{tot}\approx\sum_{n=1}^{3}P_n\sigma_n/\sigma(e^+e^-\ar b\bar b Z^0Z^0)
\approx0.59$. Since adopting one or the other of the two values
$P_1$ and $P_{tot}$
would not change the conclusions (see later on), as a first approximation we
forget the complications due to
possible $b\bar b$--decays of the on--shell $Z^0$'s in $b\bar bZ^0Z^0$ events,
and we continue to treat these latter ``inclusively''\footnote{Also,
throughout the analysis we implicitly assume that we are always considering
the right ``$b\bar b$''--pair (i.e., the tagged $b(\bar b)$
with the un--tagged $\bar b
(b)$ coming from the same $Z^0$),
this is due to the underlying cut in $M_{b\bar b}$,
which drastically suppresses (because of the narrowness of the
$Z^0$--resonance)
any contribution coming from wrong $b(\bar b)$--$jet$ combinations, with
the $jet$ eventually coming from $Z^0$--decays (as done in
ref.~\cite{eezh}).}.\par

\subsection*{Results}

Our results are presented throughout figs. 2--7, and in tabs. I--IV.\par
In figs.~2--3 we show the differential distribution
$d\sigma/dM_{Z^0Z^0}$ for $e^+e^-\rightarrow b\bar b Z^0Z^0$
events, obtained from the full matrix element (i.e., summed over all the six
contributions ${\cal M}_i^2$),
for two different values of the {\tt c.m.} energy of a
NLC, and for the same choice of Higgs masses adopted in
ref.~\cite{eezh} (see figs.~3--4 there)\footnote{Only the value
$M_{H^0}=170$ GeV, there considered at $\sqrt s=350$ GeV, has been
dropped here, since
this case would correspond to a below threshold decay $H^0\ar Z^{0*}Z^{0*}
\ar f\bar f f'\bar f'$ (where $f^{(')}$ stands for a lepton $\ell$ or
$\nu_\ell$, with $\ell=e,\mu,\tau$, or a light quark $q=u,d,s,c,b$),
with a six particle signature $(b\bar b) (f\bar f) (f'\bar f')$, which deserves
a more complicated treatment than of the one
we are interested in performing here.}. As in ref.~\cite{eezh},
in order to disentangle the signal $Z^0H^0\ar (b\bar b)(Z^0Z^0)$ from the
irreducible background we have imposed a cut around
the $Z^0$ mass, requiring that $|M_{Z^0}-M_{b\bar b}|<10$ GeV. Also,
since we are looking for events that have to be tagged by microvertex
detectors, we selected only configurations with $|\cos\theta_{b\bar b}|<0.8$
\cite{tag,GHS}.\par
Both in fig.~2 and in fig.~3 the $H^0\ar Z^0Z^0$ peaks
appear clearly visible over the flat structure of the
irreducible background, which (looking at the integrals
of the various components
(\ref{m1})--(\ref{m4}) of the matrix element) appears to be dominated
by the $H^0\ar b\bar bZ^0$ (i.e., ${\cal M}_{1}^2$)
and the $Z^0\ar b\bar b$  (i.e., ${\cal M}_{2}^2$) contributions,
which, obviously, largely pass the cut in $M_{b\bar b}$.
Moreover, the cross section corresponding to ${\cal M}_1^2$
is roughly equal to twice the signal (i.e., the integral of
${\cal M}_4^2$)
since these two processes can be approximated in terms of a
{\it production$\times$decay} reaction
$e^+e^-\ar Z^0H^0\ar Z^0(Z^0Z^0)\times BR(Z^0\ar b\bar b)$,
with the $Z^0\ar b\bar b$ decay corresponding,
in one case, to a $Z^0$ directly coming from the two--body
Bjorken process (diagram $\#17$) and, in the other case (the contribution to
this resonance coming from the $H^0\ar b\bar b$ decay followed
by a $Z^0$--bremsstrahlung off $b$--lines is in fact negligible), to a $Z^0$
from the $H^0\ar Z^0Z^0$ decay (diagrams $\#18$), and with differences (only
a few fractions of picobarns for the integrated ``cross sections'')
coming from the different kinematics of the decaying $Z^0$'s\footnote{In
the following  we will speak of a ``prompt $Z^0$'' for the
case ${\cal M}_4^2$ and of a ``$H^0$--decay $Z^0$'' for ${\cal M}_1^2$. Also we
will write
``bremsstrahlung $Z^0$'' when we will intend to indicate a $Z^0$
produced via the diagrams entering in ${\cal M}_2^2$.}.
The factor of two comes from having two $Z^0$'s in the
${\cal M}_{1}^2$
contribution that can both decay to a $b\bar b$--pair.
In the case of ${\cal M}_2^2$ we have three $Z^0$'s produced via
bremsstrahlung off the $e^+e^-$ fermion line, with one of them decaying
to the $b\bar b$--pair\footnote{We wonder
if the case ${\cal M}_{1}^2$ has to be really considered as a background,
since it includes a Higgs produced via the Bjorken reaction,
even though not peaking in the missing mass spectrum:
in fact not all the particles entering in the missing mass
come from the $H^0$. By the way, its spectrum in this
variable is quite flat and completely useless in disentangle
$H^0$--signals with respect the other backgrounds.}.\par
In tab.~I we present
the expected number of signal ($S$) and background ($B$) events together with
the statistical significance $S/\sqrt B$,
for ${\cal L} = 10 \; {\mathrm {fb}}^{-1}$ and
$\epsilon_b=1/3$, in a
window of 10 GeV around the adopted values of the Higgs mass $M_{H^0}$,
at $\sqrt s=350$ and $500$ GeV, after the cuts in
$M_{b\bar b}$ and $\cos\theta_{b\bar b}$ discussed above.
Looking at the ratio $S/\sqrt B$ it would seem that, even though with
a small number of events in some instances,
the signal is detectable. However, we have to remember that
the final goal is to look at the spectrum in missing mass and
at the total number of events when the rates for both
signal and background of both the processes \eebbzz\ and \eebbww\ are added
together in an inclusive analysis.
For that, we have plotted in figs.~4--5 the differential distribution
$d\sigma/dM_{VV}$, which is the sum of the
corresponding histograms of the two above
processes (when $VV=Z^0Z^0$ and $W^\pm W^\mp$), for the usual combination
of Higgs masses and {\tt c.m.} energies
(i.e., we sum the distributions in fig.~3(4) of \cite{eezh}
and in fig.~2(3) of this study). Then we have again integrated these
curves in a window of 10 GeV around $M_{H^0}$,
obtaining the total number of signal and background events (now picked
out of the inclusive missing mass
spectrum) and the corresponding significances shown in tab.~II\footnote{Since
in \cite{eezh} only the
significances for the cases $\sqrt s = 500$ GeV and $M_{H^0}=250$ and $300$
GeV were given for the value of $top$ mass here adopted, we list now the
remaining ones: they are 8.50(8.36)[6.20] for
$\sqrt s =350$ GeV and $M_{H^0}=185(210)[240]$ GeV, and
17.57 for $\sqrt s = 500$ GeV and $M_{H^0}=210$ GeV.}.\par
{}From figs.~4--5 and tab.~II it is then clear that adding
together the missing mass spectra of the two processes
increases the total significances, to
$\approx18(14)[20]\%$ for $\sqrt s =350$ GeV and $M_{H^0}=185(210)[240]$ GeV
and to $\approx35(36)[43]\%$ for $\sqrt s =500$ GeV and $M_{H^0}=210(250)[300]$
GeV,
with respect to those ones obtained for the process \eebbww\ only \cite{eezh}.
Now, with the values of tab.~II the only signal that
still appears quite difficult to disentangle from the irreducible background
is $M_{H^0}=300$ GeV at $\sqrt s =500$ GeV, this is also due to the fact
that for this value of $M_{H^0}$ the Higgs width is
sizably large ($\Gamma_{H^0}\approx 8.5$ GeV) and comparable
to the one of the window in $M_{VV}$
we integrate over (whereas this does not happen for the
other cases, since for them we always have $\Gamma_{H^0}<4.1$ GeV,
the value of $\Gamma_{H^0}$ for $M_{H^0}=250$ GeV).\par
Of course, at this point we could decide to integrate over
a larger window, retaining then more signal, but for
$\sqrt s =500$ GeV and (let us say) $|M_{VV}-300~{\mathrm {GeV}}|\Ord 10$ GeV
we would include also the region (around $M_{VV}\approx 310$ GeV)
where the background from \eebbww\
is maximum (compare with fig.~4 of \cite{eezh}). Therefore, this is not the
best way to proceed, and in fact
in ref.~\cite{eezh} it has instead been decided to apply cuts based on the
kinematics of \ttb--production and decay: i.e., we required
that one of the $W^\pm$'s (let us say $W^+$) failed in reproducing the
kinematics
of the \ttb--final state when coupled with either of the two $b$'s, namely that
$m_t-10~{\mathrm{GeV}}>|M_{W^+b(W^+\bar b)}|>m_t+10~{\mathrm{GeV}}$
and $E_{beam}-10~{\mathrm{GeV}}>|E_{W^+}+E_{b(\bar b)}|>
E_{beam}+10~{\mathrm{GeV}}$. But, even though these selection criteria
are quite convenient \cite{eezh}, they require
the decay products of the $W^\pm$ to be tagged: that is,
a further experimental detection effort is needed compared to the missing
mass analysis which only requires tagging the $b\bar b$--system.\par
The effect of another additional cut can then be exploited.
If we look at the spectrum
in energy $E_{b\bar b}$ of the $b\bar b$--pair, this (due to the $Z^0H^0$
two--body
kinematics of the Bjorken production) is ``practically'' mono--energetic
for a pair coming from a prompt $Z^0$, whereas it appears
quite broad if the pair is produced by a $H^0$--decay or a bremsstrahlung
$Z^0$ (figs.~6--7)\footnote{Concerning the case
of bremsstrahlung $Z^0$'s it has to be remembered (see eq.~(\ref{m1}))
that ${\cal M}_2^2$ includes also the interference of $Z^0\ar b\bar b$
with the signal $Z^0H^0\ar (b\bar b) (Z^0Z^0)$, whose effects
appear clearly visible in the ``half'' small peak below the signal one,
and which, at the end, slightly enhance the contribution of this
background.}.
Therefore, retaining only events in an appropriate window around the maximum in
$E_{b\bar b}$ could further reduce the two $Z^0$--resonant backgrounds
(and not those only) with respect to the signal.
Some care has to be taken in exploiting this possibility. In fact,
the above spectrum is really mono--energetic
only apart from photon bremsstrahlungs off $e^+e^-$--lines\footnote{Even though
photon emission
can happen also off $b\bar b$--lines, however this latter can easily
be included in the invariant mass reconstructing the $Z^0$--peak. So,
we do not stress this case further, here.}.
When such photons (namely Initial State Radiation, ISR) are
included in the computation, the energy flowing in the first $Z^0$--propagator
of diagram 17 is not a constant any longer. Therefore, the prompt $Z^0$ spectra
would appear broader than those ones plotted here.
Nevertheless, since the mean $e^+e^-$ {\tt c.m.} energy loss $\delta_{\sqrt s}$
due to ISR is, e.g.,
$\approx5\%$ at $\sqrt s=500$ GeV \cite{ISR}, one can choose a window wide
enough ($\approx \delta_{\sqrt s}\times\sqrt s$) to prevent
complications
due to such effects\footnote{The inclusion
of Linac energy spread and beamsstrahlung should not drastically
change this strategy, at least for the ``narrow'' D--D and TESLA collider
designs
(see ref.~\cite{ISR}).}.
The effectiveness of this cut is clear from tab.~III,
which presents the percentage of configurations that
give an energy of the $b\bar b$--pair in
a window of 25 GeV around the peak $E^{max}_{b\bar b}$
for the above three $b\bar b Z^0Z^0$
sub--processes.\par
So, finally, requiring for the \eebbvv\ events to have energy of
the $b\bar b$--pair in the above window around the maxima
(which are $\approx138(124)[105]$ GeV for $\sqrt s=350$ GeV
and $M_{H^0}=185(210)[240]$ GeV, and $\approx214(196)[168]$ GeV
for $\sqrt s=500$ GeV
and $M_{H^0}=210(250)[300]$ GeV for both the cases\footnote{We do not
reproduce here the figures for \eebbww, since they do not differ too much
from figs.~6--7: there the signal (various backgrounds) is(are)
as narrow (broad with respect to the signal)
as that (those) one(s) of \eebbzz.} $VV=Z^0Z^0,W^+W^-$)
and calculating the corresponding percentage of events passing this cut
(now for all the components
of both the processes \eebbww\ and \eebbzz) leads to the final number
of signal and background events, and their statistical significances,
given in tab.~IV. From which we deduce that an additional simple
cut in $E_{b\bar b}$ increases the ratios $S/\sqrt B$ up to values such
that Higgs detection should now be feasible everywhere just
by adopting a pure missing mass
analysis (i.e., without resorting to any identification of the decay products
of the vector bosons).

\subsection*{Conclusions}

In summary, in this letter we studied the production of a heavy
Higgs (with $2M_{W^\pm}<M_{H^0}<2m_t$, where $m_t=175$ GeV)
and a $Z^0$ through the Bjorken bremsstrahlung reaction
$e^+e^-\ar Z^0H^0$ at NLC energies, assuming $H^0\ar Z^0Z^0$ and
$Z^0\ar b\bar b$ and requiring a single $b$--tagging for the
$Z^0$--detection. We have also studied all the irreducible background
in \eebbzz\ events.
We found that Higgs signals, which would be clearly detectable for
\eebbzz\ on their own, still remain once we add (as needed for
the missing mass analysis) this process to \eebbww, where
$Z^0H^0\ar (b\bar b)(W^+W^-)$
and which includes among the irreducible background the huge
$t\bar t\ar b\bar b W^+W^-$ production and decay.
This was done only by imposing the following cuts on the $b\bar b$--system:
$|\cos\theta_{b\bar b}|<0.8$, $|M_{b\bar b}-M_{Z^0}|<10$ GeV and
$|E_{b\bar b}-E_{b\bar b}^{max}|<12.5$ GeV, where
$E_{b\bar b}^{max}$ is the maximum in the energy spectrum
of the $b\bar b$--pair, which is practically mono--energetic
for the $b\bar b$--pair coming from a $Z^0$ produced in the
two--body Bjorken reaction. In particular,
this latter cut turns out to be extremely useful in rejecting
the $t\bar t$--background in \eebbww\ events, thus avoiding
further cuts based on the $t\bar t$--kinematics, which,
although useful to the above aim, imply
tagging the decay products of one of the two $W^\pm$'s.
In fact, this latter procedure diminishes the attractiveness of the missing
mass analysis (which only requires tagging the $b\bar b$--system),
and also introduces a reduction
factor in the statistics
due to the branching ratio of the decaying $W^\pm$--boson.

\subsection*{Acknowledgements}

We are grateful to Tim Stelzer and Bas Tausk for useful suggestions
and constructive discussions, to Alessandro Ballestrero and Ezio Maina
for focusing our attention on some important aspects of the phenomenological
analysis here presented, and to Ghadir Abu Leil for a careful reading of the
manuscript.

\vfill
\newpage

\thispagestyle{empty}

\subsection*{Table Captions}

\begin{description}

\item[table~I  ] The expected number of $e^+e^-\ar b\bar b Z^0Z^0$ signal
and background events in the window $|M_{H^0}-M_{Z^0 Z^0}|<5$ GeV
and their statistical significance
at $\sqrt s=350$ GeV and $\sqrt s=500$ GeV
for a selection of Higgs masses after
the cuts:
$|M_{Z^0}-M_{b\bar b}|<10$ GeV and $|\cos\theta_{b\bar b}|<0.8$.
We assume that only one $b$--$jet$ is tagged with efficiency
$\epsilon_b=1/3$. The luminosity is taken to be
${\cal L} = 10~{\mathrm{fb}}^{-1}$.

\item[table~II ] The expected number of signal
and background events for $e^+e^-\ar b\bar b Z^0Z^0$ and
$e^+e^-\ar b\bar b W^+W^-$ processes, added together,
in the window $|M_{H^0}-M_{VV}|<5$ GeV
and their statistical significance
at $\sqrt s=350$ GeV and $\sqrt s=500$ GeV
for a selection of Higgs masses after
the cuts:
$|M_{Z^0}-M_{b\bar b}|<10$ GeV and $|\cos\theta_{b\bar b}|<0.8$.
We assume that only one $b$--$jet$ is tagged with efficiency
$\epsilon_b=1/3$. The luminosity is taken to be
${\cal L} = 10~{\mathrm{fb}}^{-1}$.
Numbers corresponding to the contribution of
$e^+e^-\ar b\bar b W^+W^-$ events are taken from ref.~\cite{eezh}
(assuming $m_t=175$ GeV).

\item[table~III ] Percentage of events with energy of
the $b\bar b$--pair $E_{b\bar b}$ in the window
$|E^{max}_{b\bar b}-E_{b\bar b}|<12.5$ GeV for the cases
of a prompt $Z^0$, a $H^0$--decay $Z^0$ and a bremsstrahlung
$Z^0$ (see in the text)
at $\sqrt s=350$ and $\sqrt s=500$ GeV for a selection of Higgs masses after
the cuts: $|M_{Z^0}-M_{b\bar b}|<10$ GeV and $|\cos\theta_{b\bar b}|<0.8$.

\item[table~IV  ] The expected number of signal
and background events for $e^+e^-\ar b\bar b Z^0Z^0$ and
$e^+e^-\ar b\bar b W^+W^-$ processes, added together,
in the window $|M_{H^0}-M_{VV}|<5$ GeV
and their statistical significance
at $\sqrt s=350$ GeV and $\sqrt s=500$ GeV
for a selection of Higgs masses after
the cuts:
$|M_{Z^0}-M_{b\bar b}|<10$ GeV, $|\cos\theta_{b\bar b}|<0.8$
and $|E_{b\bar b}-E_{b\bar b}^{max}|<12.5$ GeV.
We assume that only one $b$--$jet$ is tagged with efficiency
$\epsilon_b=1/3$. The luminosity is taken to be
${\cal L} = 10~{\mathrm{fb}}^{-1}$.
Numbers corresponding to the contribution of
$e^+e^-\ar b\bar b W^+W^-$ events are computed from ref.~\cite{eezh}
(assuming $m_t=175$ GeV).

\end{description}

\vfill
\newpage
\thispagestyle{empty}

\subsection*{Figure Captions}

\begin{description}

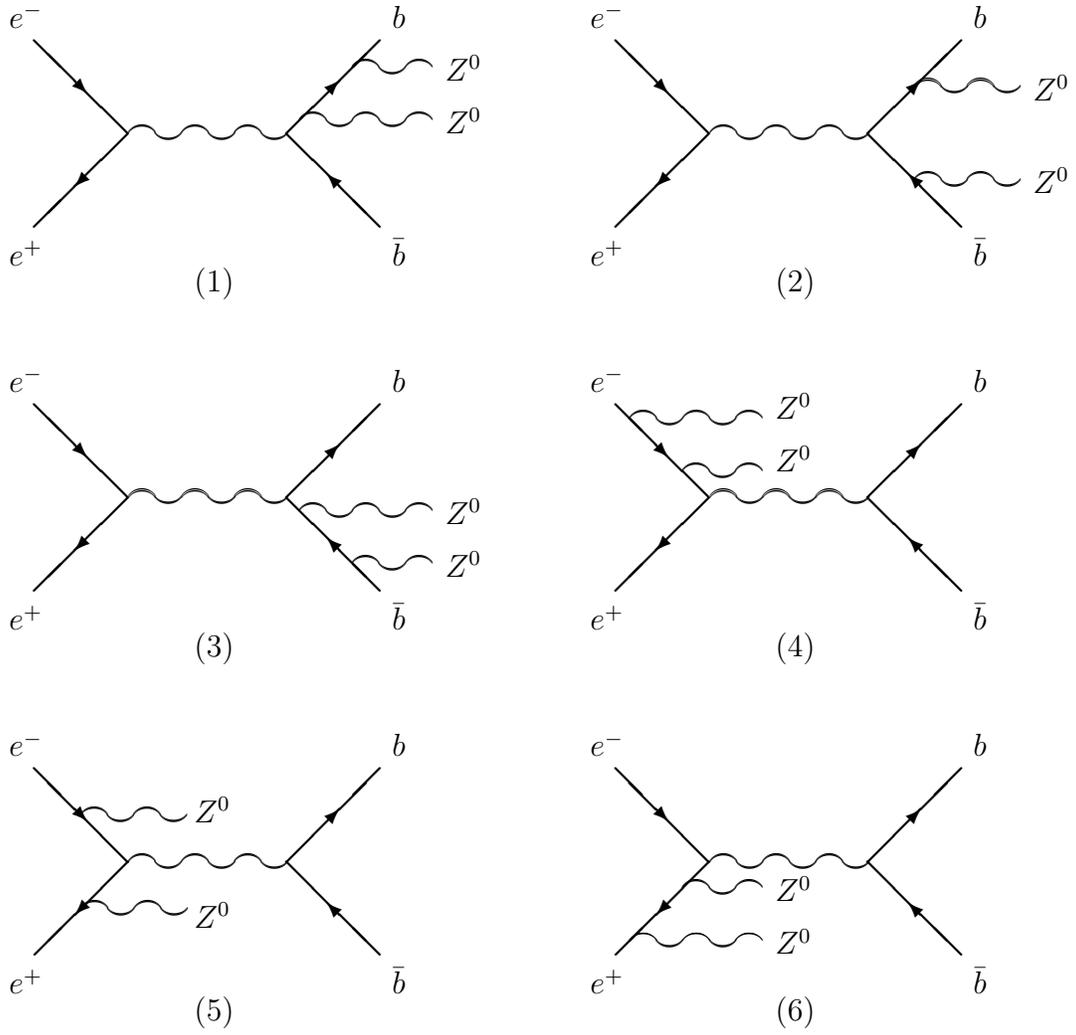
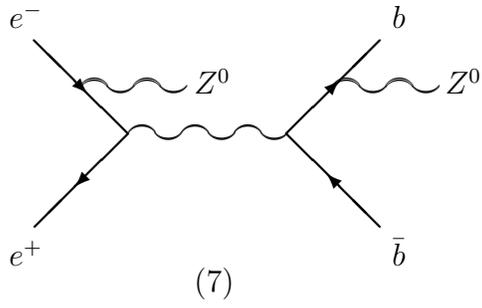
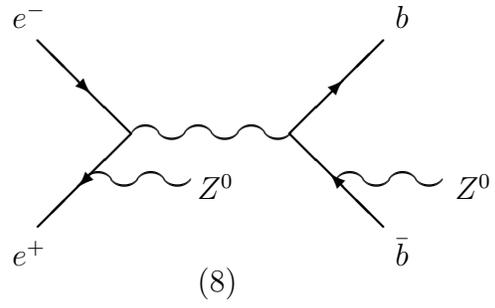
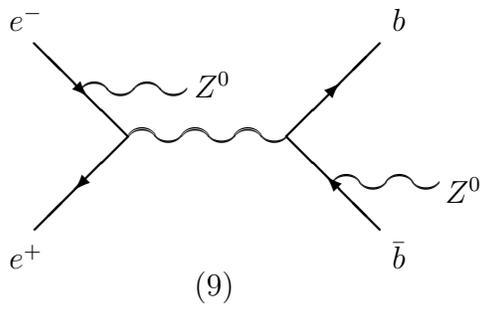
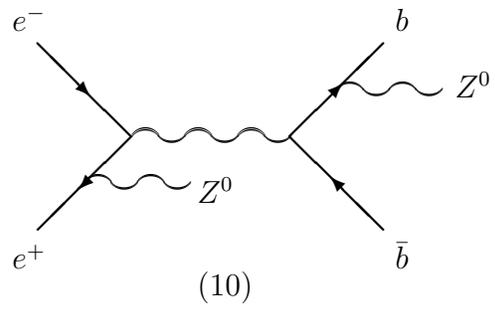
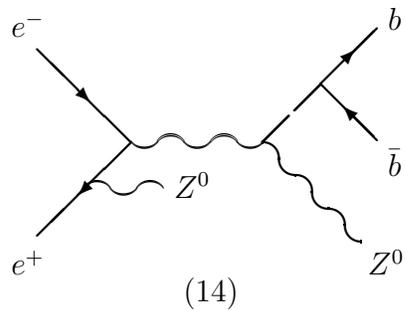
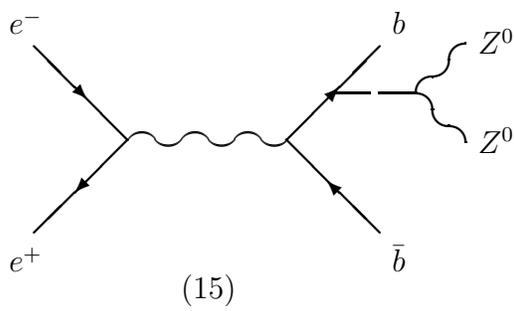
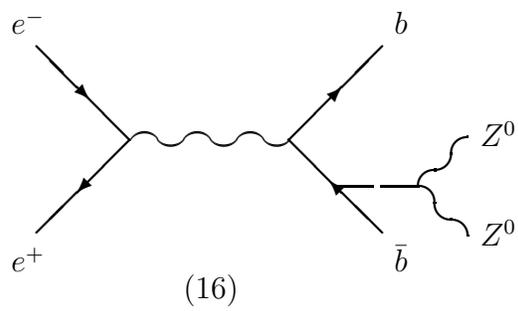
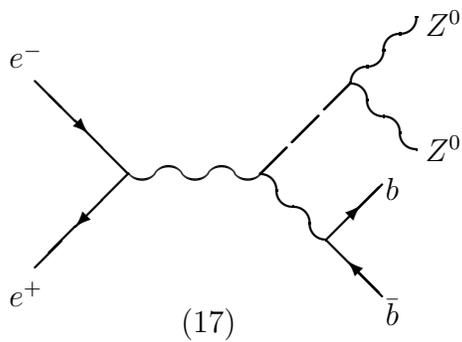
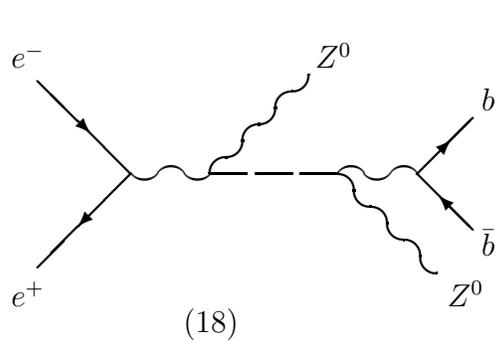
\item[figure~1 ] Feynman diagrams contributing in the lowest order to
$e^+e^-\rightarrow b\bar b Z^0Z^0$ (those ones obtainable
by exchanging the two $Z^0$ bosons are not shown).
Internal wavy lines represent a $\gamma$ or a $Z^0$,
as appropriate. Internal dashed lines represent a Higgs boson.

\item[figure~2] The differential distribution
$d\sigma/dM_{Z^0Z^0}$ for
$e^+e^-\rightarrow b\bar b Z^0Z^0$ (full matrix element
with all Higgs contributions),
at $\sqrt s=350$ GeV, for
$M_{H^0}=185$ GeV (continuous line),
$M_{H^0}=210$ GeV (dashed line) and
$M_{H^0}=240$ GeV (dotted line), with the following cuts:
$|M_{Z^0}-M_{b\bar b}|<10$ GeV and $|\cos\theta_{b\bar b}|<0.8$.

\item[figure~3] The differential distribution
$d\sigma/dM_{Z^0Z^0}$ for
$e^+e^-\rightarrow b\bar b Z^0Z^0$ (full matrix element
with all Higgs contributions),
at $\sqrt s=500$ GeV, for
$M_{H^0}=210$ GeV (continuous line),
$M_{H^0}=250$ GeV (dashed line) and
$M_{H^0}=300$ GeV (dotted line), with the following cuts:
$|M_{Z^0}-M_{b\bar b}|<10$ GeV and $|\cos\theta_{b\bar b}|<0.8$.

\item[figure~4] The differential distribution
$d\sigma/dM_{VV}$,
for $e^+e^-\ar b\bar b Z^0Z^0$ and
$e^+e^-\ar b\bar b W^+W^-$ processes
(full matrix elements
with all Higgs contributions), added together,
at $\sqrt s=350$ GeV, for
$M_{H^0}=185$ GeV (continuous line),
$M_{H^0}=210$ GeV (dashed line) and
$M_{H^0}=240$ GeV (dotted line), with the following cuts:
$|M_{Z^0}-M_{b\bar b}|<10$ GeV and $|\cos\theta_{b\bar b}|<0.8$.
Plots corresponding to the contribution of
$e^+e^-\ar b\bar b W^+W^-$ events are taken from ref.~\cite{eezh}
(assuming $m_t=175$ GeV).

\item[figure~5] The differential distribution
$d\sigma/dM_{VV}$,
for $e^+e^-\ar b\bar b Z^0Z^0$ and
$e^+e^-\ar b\bar b W^+W^-$ processes
(full matrix elements
with all Higgs contributions), added together,
at $\sqrt s=500$ GeV, for
$M_{H^0}=210$ GeV (continuous line),
$M_{H^0}=250$ GeV (dashed line) and
$M_{H^0}=300$ GeV (dotted line), with the following cuts:
$|M_{Z^0}-M_{b\bar b}|<10$ GeV and $|\cos\theta_{b\bar b}|<0.8$.
Plots corresponding to the contribution of
$e^+e^-\ar b\bar b W^+W^-$ events are taken from ref.~\cite{eezh}
(assuming $m_t=175$ GeV).

\item[figure~6] The differential distribution
$d\sigma/dE_{b\bar b}/\sigma$ for the signal
$e^+e^-\rightarrow Z^0H^0$ with the $b\bar b$--pair
coming from the prompt $Z^0$ (continuous line),
from a $H^0$--decay $Z^0$ (dashed line) and from
a bremsstrahlung $Z^0$ (dotted line),
at $\sqrt s=350$ GeV, for $M_{H^0}=185,210$ and 240 GeV, with the following
cuts:
$|M_{Z^0}-M_{b\bar b}|<10$ GeV and $|\cos\theta_{b\bar b}|<0.8$.

\end{description}

\vfill
\newpage
\thispagestyle{empty}

\begin{description}

\item[figure~7] The differential distribution
$d\sigma/dE_{b\bar b}/\sigma$ for the signal
$e^+e^-\rightarrow Z^0H^0$ with the $b\bar b$--pair
coming from the prompt $Z^0$ (continuous line),
from a $H^0$--decay $Z^0$ (dashed line) and from
a bremsstrahlung $Z^0$ (dotted line),
at $\sqrt s=500$ GeV, for $M_{H^0}=210,250$ and 300 GeV, with the following
cuts:
$|M_{Z^0}-M_{b\bar b}|<10$ GeV and $|\cos\theta_{b\bar b}|<0.8$.

\end{description}
\vfill

\newpage
\pagestyle{empty}

\vspace*{-1.5cm}

\begin{table}
\begin{center}
\begin{tabular}{|c|c|c|c|}
\hline
\rule[-0.5cm]{0cm}{1.1cm}
$M_{H^0}~{\mathrm{(GeV)}}$ &\makebox[2.2cm]{Signal} & Background
& $S/\sqrt{B}$  \\ \hline\hline
\multicolumn{4}{|c|}
{\rule[-0.5cm]{0cm}{1.1cm}
 $~~\sqrt s=350~{\mathrm{GeV}}$ }
\\ \hline
\rule[-0.6cm]{0cm}{1.3cm}
$185$ & $7.45$ & $0.18$ & $17.63$    \\ 
\rule[-0.6cm]{0cm}{.9cm}
$210$ & $9.00$ & $3.64$ & $4.72$     \\  
\rule[-0.6cm]{0cm}{.9cm}
$240$ & $4.44$ & $1.13$ & $4.17$     \\  \hline\hline
\multicolumn{4}{|c|}
{\rule[-0.5cm]{0cm}{1.1cm}
$~~\sqrt s=500~{\mathrm{GeV}}$ }
 \\ \hline
\rule[-0.6cm]{0cm}{1.3cm}
$210$ & $6.78$ & $0.058$ & $28.20$    \\  
\rule[-0.6cm]{0cm}{.9cm}
$250$ & $4.93$ & $0.73$ & $5.75$    \\  
\rule[-0.6cm]{0cm}{.9cm}
$300$ & $2.51$ & $0.51$ & $3.52$     \\ \hline\hline
\multicolumn{4}{|c|}
{\rule[-0.5cm]{0cm}{1.1cm}
${\cal L} = 10~{\mathrm{fb}}^{-1}~~~~\epsilon_b=1/3$ }
 \\ \hline
\multicolumn{4}{c}
{\rule{0cm}{.9cm}
{\Large Table I}}  \\
\multicolumn{4}{c}
{\rule{0cm}{.9cm}}

\end{tabular}
\end{center}
\end{table}

\vfill

\newpage
\pagestyle{empty}

\vspace*{-1.5cm}

\begin{table}
\begin{center}
\begin{tabular}{|c|c|c|c|}
\hline
\rule[-0.5cm]{0cm}{1.1cm}
$M_{H^0}~{\mathrm{(GeV)}}$ &\makebox[2.2cm]{Signals} & Backgrounds
& $S/\sqrt{B}$  \\ \hline\hline
\multicolumn{4}{|c|}
{\rule[-0.5cm]{0cm}{1.1cm}
 $~~\sqrt s=350~{\mathrm{GeV}}$ }
\\ \hline
\rule[-0.6cm]{0cm}{1.3cm}
$185$ & $47.85$ & $22.79$ & $10.02$    \\ 
\rule[-0.6cm]{0cm}{.9cm}
$210$ & $32.47$ & $11.53$ & $9.56$     \\  
\rule[-0.6cm]{0cm}{.9cm}
$240$ & $15.06$ & $4.06$ & $7.47$     \\  \hline\hline
\multicolumn{4}{|c|}
{\rule[-0.5cm]{0cm}{1.1cm}
$~~\sqrt s=500~{\mathrm{GeV}}$ }
 \\ \hline
\rule[-0.6cm]{0cm}{1.3cm}
$210$ & $24.48$ & $1.07$ & $23.65$    \\  
\rule[-0.6cm]{0cm}{.9cm}
$250$ & $16.56$ & $9.01$ & $5.52$    \\  
\rule[-0.6cm]{0cm}{.9cm}
$300$ & $8.17$ & $17.92$ & $1.93$     \\ \hline\hline
\multicolumn{4}{|c|}
{\rule[-0.5cm]{0cm}{1.1cm}
${\cal L} = 10~{\mathrm{fb}}^{-1}~~~~\epsilon_b=1/3$ }
 \\ \hline
\multicolumn{4}{c}
{\rule{0cm}{.9cm}
{\Large Table II}}  \\
\multicolumn{4}{c}
{\rule{0cm}{.9cm}}

\end{tabular}
\end{center}
\end{table}

\vfill
\newpage
\pagestyle{empty}

\vspace*{-1.5cm}

\begin{table}
\begin{center}
\begin{tabular}{|c|c|c|c|}
\hline
\rule[-0.5cm]{0cm}{1.1cm}
$M_{H^0}~{\mathrm{(GeV)}}$ &\makebox[2.2cm]{Prompt $Z^0$} & $H^0$--decay $Z^0$
& Bremsstrahlung $Z^0$  \\ \hline\hline
\multicolumn{4}{|c|}
{\rule[-0.5cm]{0cm}{1.1cm}
 $~~\sqrt s=350~{\mathrm{GeV}}$ }
\\ \hline
\rule[-0.6cm]{0cm}{1.3cm}
$185$ & $98\%$ & $0.63\%$ & $27\%$    \\ 
\rule[-0.6cm]{0cm}{.9cm}
$210$ & $99\%$ & $56\%$ & $70\%$     \\  
\rule[-0.6cm]{0cm}{.9cm}
$240$ & $98\%$ & $36\%$ & $61\%$     \\  \hline\hline
\multicolumn{4}{|c|}
{\rule[-0.5cm]{0cm}{1.1cm}
$~~\sqrt s=500~{\mathrm{GeV}}$ }
 \\ \hline
\rule[-0.6cm]{0cm}{1.3cm}
$210$ & $98\%$ & $0.36\%$ & $18\%$    \\  
\rule[-0.6cm]{0cm}{.9cm}
$250$ & $96\%$ & $23\%$ & $23\%$    \\  
\rule[-0.6cm]{0cm}{.9cm}
$300$ & $91\%$ & $23\%$ & $22\%$     \\ \hline\hline
\multicolumn{4}{|c|}
{\rule[-0.5cm]{0cm}{1.1cm}
$Z^0\ar b\bar b$}
 \\ \hline
\multicolumn{4}{c}
{\rule{0cm}{.9cm}
{\Large Table III}}  \\
\multicolumn{4}{c}
{\rule{0cm}{.9cm}}

\end{tabular}
\end{center}
\end{table}

\vfill

\newpage
\pagestyle{empty}

\vspace*{-1.5cm}

\begin{table}
\begin{center}
\begin{tabular}{|c|c|c|c|}
\hline
\rule[-0.5cm]{0cm}{1.1cm}
$M_{H^0}~{\mathrm{(GeV)}}$ &\makebox[2.2cm]{Signals} & Backgrounds
& $S/\sqrt{B}$  \\ \hline\hline
\multicolumn{4}{|c|}
{\rule[-0.5cm]{0cm}{1.1cm}
 $~~\sqrt s=350~{\mathrm{GeV}}$ }
\\ \hline
\rule[-0.6cm]{0cm}{1.3cm}
$185$ & $47.43$ & $18.60$ & $11.00$    \\ 
\rule[-0.6cm]{0cm}{.9cm}
$210$ & $32.08$ & $5.97$ & $13.13$     \\  
\rule[-0.6cm]{0cm}{.9cm}
$240$ & $14.79$ & $1.40$ & $12.50$     \\  \hline\hline
\multicolumn{4}{|c|}
{\rule[-0.5cm]{0cm}{1.1cm}
$~~\sqrt s=500~{\mathrm{GeV}}$ }
 \\ \hline
\rule[-0.6cm]{0cm}{1.3cm}
$210$ & $23.91$ & $0.14$ & $62.84$    \\  
\rule[-0.6cm]{0cm}{.9cm}
$250$ & $15.87$ & $1.54$ & $12.79$    \\  
\rule[-0.6cm]{0cm}{.9cm}
$300$ & $7.42$ & $5.09$ & $3.29$     \\ \hline\hline
\multicolumn{4}{|c|}
{\rule[-0.5cm]{0cm}{1.1cm}
${\cal L} = 10~{\mathrm{fb}}^{-1}~~~~\epsilon_b=1/3$ }
 \\ \hline
\multicolumn{4}{c}
{\rule{0cm}{.9cm}
{\Large Table IV}}  \\
\multicolumn{4}{c}
{\rule{0cm}{.9cm}}

\end{tabular}
\end{center}
\end{table}

\vfill
\newpage
\pagestyle{empty}
\
\vskip 2.0cm

\begin{picture}(10000,8000)
\THICKLINES
\bigphotons
\drawline\photon[\W\REG](10000,8000)[6]
\drawline\fermion[\NW\REG](\photonbackx,\photonbacky)[5000]
\drawarrow[\SE\ATBASE](\pmidx,\pmidy)
\drawline\fermion[\SW\REG](\photonbackx,\photonbacky)[5000]
\drawarrow[\SW\ATBASE](\pmidx,\pmidy)
\drawline\fermion[\NE\REG](\photonfrontx,\photonfronty)[5000]
\drawarrow[\NE\ATBASE](\pmidx,\pmidy)
\drawline\photon[\E\REG](12500,10500)[3]
\drawline\photon[\E\REG](10500,8500)[5]
\drawline\fermion[\SE\REG](10000,8000)[5000]
\drawarrow[\NW\ATBASE](\pmidx,\pmidy)
\put(-500,12000){$e^-$}
\put(-500,3000){$e^+$}
\put(14000,12000){$b$}
\put(14000,3000){$\bar b$}
\put(16000,10000){$Z^0$}
\put(16000,8000){$Z^0$}
\put(6500,2000){$(1)$}
\drawline\photon[\W\REG](32000,8000)[6]
\drawline\fermion[\NW\REG](\photonbackx,\photonbacky)[5000]
\drawarrow[\SE\ATBASE](\pmidx,\pmidy)
\drawline\fermion[\SW\REG](\photonbackx,\photonbacky)[5000]
\drawarrow[\SW\ATBASE](\pmidx,\pmidy)
\drawline\fermion[\NE\REG](\photonfrontx,\photonfronty)[5000]
\drawarrow[\NE\ATBASE](\pmidx,\pmidy)
\drawline\photon[\E\REG](\pmidx,\pmidy)[4]
\drawline\fermion[\SE\REG](32000,8000)[5000]
\drawarrow[\NW\ATBASE](\pmidx,\pmidy)
\drawline\photon[\E\REG](\pmidx,\pmidy)[4]
\put(21500,12000){$e^-$}
\put(21500,3000){$e^+$}
\put(36000,12000){$b$}
\put(36000,3000){$\bar b$}
\put(38250,9250){$Z^0$}
\put(38250,5750){$Z^0$}
\put(28500,2000){$(2)$}
\end{picture}

\vskip 2.0cm

\begin{picture}(10000,8000)
\THICKLINES
\bigphotons
\drawline\photon[\W\REG](10000,8000)[6]
\drawline\fermion[\NW\REG](\photonbackx,\photonbacky)[5000]
\drawarrow[\SE\ATBASE](\pmidx,\pmidy)
\drawline\fermion[\SW\REG](\photonbackx,\photonbacky)[5000]
\drawarrow[\SW\ATBASE](\pmidx,\pmidy)
\drawline\fermion[\NE\REG](\photonfrontx,\photonfronty)[5000]
\drawarrow[\NE\ATBASE](\pmidx,\pmidy)
\drawline\fermion[\SE\REG](10000,8000)[5000]
\drawarrow[\NW\ATBASE](\pmidx,\pmidy)
\drawline\photon[\E\REG](12500,5500)[3]
\drawline\photon[\E\REG](10500,7500)[5]
\put(-500,12000){$e^-$}
\put(-500,3000){$e^+$}
\put(14000,12000){$b$}
\put(14000,3000){$\bar b$}
\put(16000,5000){$Z^0$}
\put(16000,7000){$Z^0$}
\put(6500,2000){$(3)$}
\drawline\photon[\W\REG](32000,8000)[6]
\drawline\fermion[\NW\REG](\photonbackx,\photonbacky)[5000]
\drawarrow[\SE\ATBASE](\pmidx,\pmidy)
\drawline\fermion[\SW\REG](\photonbackx,\photonbacky)[5000]
\drawarrow[\SW\ATBASE](\pmidx,\pmidy)
\drawline\photon[\E\REG](23000,11000)[5]
\drawline\photon[\E\REG](25000,9000)[3]
\drawline\fermion[\NE\REG](32000,8000)[5000]
\drawarrow[\NE\ATBASE](\pmidx,\pmidy)
\drawline\fermion[\SE\REG](32000,8000)[5000]
\drawarrow[\NW\ATBASE](\pmidx,\pmidy)
\put(21500,12000){$e^-$}
\put(21500,3000){$e^+$}
\put(36000,12000){$b$}
\put(36000,3000){$\bar b$}
\put(28500,11000){$Z^0$}
\put(28500,9000){$Z^0$}
\put(28500,2000){$(4)$}
\end{picture}

\vskip 2.0cm

\begin{picture}(10000,8000)
\THICKLINES
\bigphotons
\drawline\photon[\W\REG](10000,8000)[6]
\drawline\fermion[\NW\REG](\photonbackx,\photonbacky)[5000]
\drawarrow[\SE\ATBASE](\pmidx,\pmidy)
\drawline\photon[\E\REG](\pmidx,\pmidy)[4]
\drawline\fermion[\SW\REG](4000,8000)[5000]
\drawarrow[\SW\ATBASE](\pmidx,\pmidy)
\drawline\photon[\E\REG](\pmidx,\pmidy)[4]
\drawline\fermion[\NE\REG](10000,8000)[5000]
\drawarrow[\NE\ATBASE](\pmidx,\pmidy)
\drawline\fermion[\SE\REG](10000,8000)[5000]
\drawarrow[\NW\ATBASE](\pmidx,\pmidy)
\put(-500,12000){$e^-$}
\put(-500,3000){$e^+$}
\put(14000,12000){$b$}
\put(14000,3000){$\bar b$}
\put(6500,9500){$Z^0$}
\put(6500,5500){$Z^0$}
\put(6500,2000){$(5)$}
\drawline\photon[\W\REG](32000,8000)[6]
\drawline\fermion[\NW\REG](\photonbackx,\photonbacky)[5000]
\drawarrow[\SE\ATBASE](\pmidx,\pmidy)
\drawline\photon[\E\REG](23000,5000)[5]
\drawline\photon[\E\REG](25000,7000)[3]
\drawline\fermion[\SW\REG](26000,8000)[5000]
\drawarrow[\SW\ATBASE](\pmidx,\pmidy)
\drawline\fermion[\NE\REG](32000,8000)[5000]
\drawarrow[\NE\ATBASE](\pmidx,\pmidy)
\drawline\fermion[\SE\REG](32000,8000)[5000]
\drawarrow[\NW\ATBASE](\pmidx,\pmidy)
\put(21500,12000){$e^-$}
\put(21500,3000){$e^+$}
\put(36000,12000){$b$}
\put(36000,3000){$\bar b$}
\put(28500,4500){$Z^0$}
\put(28500,6500){$Z^0$}
\put(28500,2000){$(6)$}
\end{picture}

\vskip 1.0cm
\centerline{\bf\Large Figure 1}

\vfill
\newpage
\thispagestyle{empty}
\
\vskip 2.0cm

\begin{picture}(10000,8000)
\THICKLINES
\bigphotons
\drawline\photon[\W\REG](10000,8000)[6]
\drawline\fermion[\NW\REG](\photonbackx,\photonbacky)[5000]
\drawarrow[\SE\ATBASE](\pmidx,\pmidy)
\drawline\photon[\E\REG](\pmidx,\pmidy)[4]
\drawline\fermion[\SW\REG](4000,8000)[5000]
\drawarrow[\SW\ATBASE](\pmidx,\pmidy)
\drawline\fermion[\NE\REG](10000,8000)[5000]
\drawarrow[\NE\ATBASE](\pmidx,\pmidy)
\drawline\photon[\E\REG](\pmidx,\pmidy)[4]
\drawline\fermion[\SE\REG](10000,8000)[5000]
\drawarrow[\NW\ATBASE](\pmidx,\pmidy)
\put(-500,12000){$e^-$}
\put(-500,3000){$e^+$}
\put(14000,12000){$b$}
\put(14000,3000){$\bar b$}
\put(16000,9500){$Z^0$}
\put(6500,9500){$Z^0$}
\put(6500,2000){$(7)$}
\drawline\photon[\W\REG](32000,8000)[6]
\drawline\fermion[\NW\REG](\photonbackx,\photonbacky)[5000]
\drawarrow[\SE\ATBASE](\pmidx,\pmidy)
\drawline\fermion[\SW\REG](26000,8000)[5000]
\drawarrow[\SW\ATBASE](\pmidx,\pmidy)
\drawline\photon[\E\REG](\pmidx,\pmidy)[4]
\drawline\fermion[\NE\REG](32000,8000)[5000]
\drawarrow[\NE\ATBASE](\pmidx,\pmidy)
\drawline\fermion[\SE\REG](32000,8000)[5000]
\drawarrow[\NW\ATBASE](\pmidx,\pmidy)
\drawline\photon[\E\REG](\pmidx,\pmidy)[4]
\put(21500,12000){$e^-$}
\put(21500,3000){$e^+$}
\put(36000,12000){$b$}
\put(36000,3000){$\bar b$}
\put(38250,5500){$Z^0$}
\put(28500,5500){$Z^0$}
\put(28500,2000){$(8)$}
\end{picture}

\vskip 2.0cm

\begin{picture}(10000,8000)
\THICKLINES
\bigphotons
\drawline\photon[\W\REG](10000,8000)[6]
\drawline\fermion[\NW\REG](\photonbackx,\photonbacky)[5000]
\drawarrow[\SE\ATBASE](\pmidx,\pmidy)
\drawline\photon[\E\REG](\pmidx,\pmidy)[4]
\drawline\fermion[\SW\REG](4000,8000)[5000]
\drawarrow[\SW\ATBASE](\pmidx,\pmidy)
\drawline\fermion[\NE\REG](10000,8000)[5000]
\drawarrow[\NE\ATBASE](\pmidx,\pmidy)
\drawline\fermion[\SE\REG](10000,8000)[5000]
\drawarrow[\NW\ATBASE](\pmidx,\pmidy)
\drawline\photon[\E\REG](\pmidx,\pmidy)[4]
\put(-500,12000){$e^-$}
\put(-500,3000){$e^+$}
\put(14000,12000){$b$}
\put(14000,3000){$\bar b$}
\put(16000,5500){$Z^0$}
\put(6500,9500){$Z^0$}
\put(6500,2000){$(9)$}
\drawline\photon[\W\REG](32000,8000)[6]
\drawline\fermion[\NW\REG](\photonbackx,\photonbacky)[5000]
\drawarrow[\SE\ATBASE](\pmidx,\pmidy)
\drawline\fermion[\SW\REG](26000,8000)[5000]
\drawarrow[\SW\ATBASE](\pmidx,\pmidy)
\drawline\photon[\E\REG](\pmidx,\pmidy)[4]
\drawline\fermion[\NE\REG](32000,8000)[5000]
\drawarrow[\NE\ATBASE](\pmidx,\pmidy)
\drawline\photon[\E\REG](\pmidx,\pmidy)[4]
\drawline\fermion[\SE\REG](32000,8000)[5000]
\drawarrow[\NW\ATBASE](\pmidx,\pmidy)
\put(21500,12000){$e^-$}
\put(21500,3000){$e^+$}
\put(36000,12000){$b$}
\put(36000,3000){$\bar b$}
\put(38250,9500){$Z^0$}
\put(28500,5500){$Z^0$}
\put(28500,2000){$(10)$}
\end{picture}

\vskip 1.0cm
\centerline{\bf\Large Figure 1 (Continued)}

\vfill
\newpage
\pagestyle{empty}
\
\vskip 2.0cm

\begin{picture}(10000,8000)
\THICKLINES
\bigphotons
\drawline\photon[\W\REG](9000,8000)[5]
\drawline\fermion[\NW\REG](\photonbackx,\photonbacky)[5000]
\drawarrow[\SE\ATBASE](\pmidx,\pmidy)
\drawline\fermion[\SW\REG](\fermionfrontx,\fermionfronty)[5000]
\drawarrow[\SW\ATBASE](\pmidx,\pmidy)
\drawline\photon[\NE\REG](9000,8000)[6]
\seglength=1416  \gaplength=300  
\drawline\scalar[\SE\REG](\photonfrontx,\photonfronty)[2]
\drawline\fermion[\NE\REG](\scalarbackx,\scalarbacky)[3000]
\drawarrow[\NE\ATBASE](\pmidx,\pmidy)
\drawline\photon[\E\REG](\pmidx,\pmidy)[3]
\drawline\fermion[\SE\REG](\scalarbackx,\scalarbacky)[3000]
\drawarrow[\NW\ATBASE](\pmidx,\pmidy)
\put(-500,12000){$e^-$}
\put(-500,3000){$e^+$}
\put(13000,12000){$Z^0$}
\put(15500,6500){$Z^0$}
\put(13850,8000){$b$}
\put(13850,2750){$\bar b$}
\put(6000,2000){$(11)$}
\drawline\photon[\W\REG](31000,8000)[5]
\drawline\fermion[\NW\REG](\photonbackx,\photonbacky)[5000]
\drawarrow[\SE\ATBASE](\pmidx,\pmidy)
\drawline\fermion[\SW\REG](\fermionfrontx,\fermionfronty)[5000]
\drawarrow[\SW\ATBASE](\pmidx,\pmidy)
\drawline\photon[\SE\REG](31000,8000)[6]
\seglength=1416  \gaplength=300  
\drawline\scalar[\NE\REG](\photonfrontx,\photonfronty)[2]
\drawline\fermion[\NE\REG](\scalarbackx,\scalarbacky)[3000]
\drawarrow[\NE\ATBASE](\pmidx,\pmidy)
\drawline\fermion[\SE\REG](\scalarbackx,\scalarbacky)[3000]
\drawarrow[\NW\ATBASE](\pmidx,\pmidy)
\drawline\photon[\E\REG](\pmidx,\pmidy)[3]
\put(21500,12000){$e^-$}
\put(21500,3000){$e^+$}
\put(35000,3000){$Z^0$}
\put(37500,8750){$Z^0$}
\put(35750,12250){$b$}
\put(35750,6750){$\bar b$}
\put(28000,2000){$(12)$}
\end{picture}

\vskip 2.0cm

\begin{picture}(10000,8000)
\THICKLINES
\bigphotons
\drawline\photon[\W\REG](9000,8000)[5]
\drawline\fermion[\NW\REG](\photonbackx,\photonbacky)[5000]
\drawarrow[\SE\ATBASE](\pmidx,\pmidy)
\drawline\photon[\E\REG](\pmidx,\pmidy)[3]
\drawline\fermion[\SW\REG](\fermionfrontx,\fermionfronty)[5000]
\drawarrow[\SW\ATBASE](\pmidx,\pmidy)
\drawline\photon[\NE\REG](9000,8000)[6]
\seglength=1416  \gaplength=300  
\drawline\scalar[\SE\REG](\photonfrontx,\photonfronty)[2]
\drawline\fermion[\NE\REG](\scalarbackx,\scalarbacky)[3000]
\drawarrow[\NE\ATBASE](\pmidx,\pmidy)
\drawline\fermion[\SE\REG](\scalarbackx,\scalarbacky)[3000]
\drawarrow[\NW\ATBASE](\pmidx,\pmidy)
\put(-500,12000){$e^-$}
\put(-500,3000){$e^+$}
\put(13000,12000){$Z^0$}
\put(5650,9500){$Z^0$}
\put(13850,8000){$b$}
\put(13850,2750){$\bar b$}
\put(6000,2000){$(13)$}
\drawline\photon[\W\REG](31000,8000)[5]
\drawline\fermion[\NW\REG](\photonbackx,\photonbacky)[5000]
\drawarrow[\SE\ATBASE](\pmidx,\pmidy)
\drawline\fermion[\SW\REG](\fermionfrontx,\fermionfronty)[5000]
\drawarrow[\SW\ATBASE](\pmidx,\pmidy)
\drawline\photon[\E\REG](\pmidx,\pmidy)[3]
\drawline\photon[\SE\REG](31000,8000)[6]
\seglength=1416  \gaplength=300  
\drawline\scalar[\NE\REG](\photonfrontx,\photonfronty)[2]
\drawline\fermion[\NE\REG](\scalarbackx,\scalarbacky)[3000]
\drawarrow[\NE\ATBASE](\pmidx,\pmidy)
\drawline\fermion[\SE\REG](\scalarbackx,\scalarbacky)[3000]
\drawarrow[\NW\ATBASE](\pmidx,\pmidy)
\put(21500,12000){$e^-$}
\put(21500,3000){$e^+$}
\put(35000,3000){$Z^0$}
\put(27650,5900){$Z^0$}
\put(35750,12250){$b$}
\put(35750,6750){$\bar b$}
\put(28000,2000){$(14)$}
\end{picture}

\vskip 2.0cm

\begin{picture}(10000,8000)
\THICKLINES
\bigphotons
\drawline\photon[\W\REG](10000,8000)[6]
\drawline\fermion[\NW\REG](\photonbackx,\photonbacky)[5000]
\drawarrow[\SE\ATBASE](\pmidx,\pmidy)
\drawline\fermion[\SW\REG](\photonbackx,\photonbacky)[5000]
\drawarrow[\SW\ATBASE](\pmidx,\pmidy)
\drawline\fermion[\NE\REG](\photonfrontx,\photonfronty)[5000]
\drawarrow[\NE\ATBASE](\pmidx,\pmidy)
\seglength=1416  \gaplength=300  
\drawline\scalar[\E\REG](\pmidx,\pmidy)[2]
\drawline\photon[\NE\REG](\scalarbackx,\scalarbacky)[3]
\drawline\photon[\SE\REG](\scalarbackx,\scalarbacky)[3]
\drawline\fermion[\SE\REG](10000,8000)[5000]
\drawarrow[\NW\ATBASE](\pmidx,\pmidy)
\put(-500,12000){$e^-$}
\put(-500,3000){$e^+$}
\put(14000,12000){$b$}
\put(14000,3000){$\bar b$}
\put(17250,11250){$Z^0$}
\put(17250,7500){$Z^0$}
\put(6000,2000){$(15)$}
\drawline\photon[\W\REG](32000,8000)[6]
\drawline\fermion[\NW\REG](\photonbackx,\photonbacky)[5000]
\drawarrow[\SE\ATBASE](\pmidx,\pmidy)
\drawline\fermion[\SW\REG](\photonbackx,\photonbacky)[5000]
\drawarrow[\SW\ATBASE](\pmidx,\pmidy)
\drawline\fermion[\NE\REG](\photonfrontx,\photonfronty)[5000]
\drawarrow[\NE\ATBASE](\pmidx,\pmidy)
\drawline\fermion[\SE\REG](32000,8000)[5000]
\drawarrow[\NW\ATBASE](\pmidx,\pmidy)
\seglength=1416  \gaplength=300  
\drawline\scalar[\E\REG](\pmidx,\pmidy)[2]
\drawline\photon[\NE\REG](\scalarbackx,\scalarbacky)[3]
\drawline\photon[\SE\REG](\scalarbackx,\scalarbacky)[3]
\put(21500,12000){$e^-$}
\put(21500,3000){$e^+$}
\put(36000,12000){$b$}
\put(36000,3000){$\bar b$}
\put(39250,7750){$Z^0$}
\put(39250,4000){$Z^0$}
\put(28000,2000){$(16)$}
\end{picture}

\vskip 2.0cm

\begin{picture}(10000,8000)
\THICKLINES
\bigphotons
\drawline\photon[\W\REG](9000,8000)[5]
\drawline\fermion[\NW\REG](\photonbackx,\photonbacky)[5000]
\drawarrow[\SE\ATBASE](\pmidx,\pmidy)
\drawline\fermion[\SW\REG](\photonbackx,\photonbacky)[5000]
\drawarrow[\SW\ATBASE](\pmidx,\pmidy)
\seglength=1416  \gaplength=300  
\drawline\scalar[\NE\REG](\photonfrontx,\photonfronty)[3]
\drawline\photon[\NE\REG](\scalarbackx,\scalarbacky)[4]
\drawline\photon[\SE\REG](\scalarbackx,\scalarbacky)[4]
\drawline\photon[\SE\REG](9000,8000)[4]
\drawline\fermion[\NE\REG](\photonbackx,\photonbacky)[3000]
\drawarrow[\NE\ATBASE](\pmidx,\pmidy)
\drawline\fermion[\SE\REG](\photonbackx,\photonbacky)[3000]
\drawarrow[\NW\ATBASE](\pmidx,\pmidy)
\put(-500,12000){$e^-$}
\put(-500,3000){$e^+$}
\put(15250,13250){$Z^0$}
\put(15250,8500){$Z^0$}
\put(13750,7000){$b$}
\put(13750,2250){$\bar b$}
\put(6000,2000){$(17)$}
\drawline\photon[\W\REG](29000,8000)[3]
\drawline\fermion[\NW\REG](\photonbackx,\photonbacky)[5000]
\drawarrow[\SE\ATBASE](\pmidx,\pmidy)
\drawline\fermion[\SW\REG](\photonbackx,\photonbacky)[5000]
\drawarrow[\SW\ATBASE](\pmidx,\pmidy)
\drawline\photon[\NE\REG](\photonfrontx,\photonfronty)[6]
\seglength=1416  \gaplength=300  
\drawline\scalar[\E\REG](29000,8000)[3]
\drawline\photon[\SE\REG](\scalarbackx,\scalarbacky)[6]
\drawline\photon[\E\REG](\scalarbackx,\scalarbacky)[3]
\drawline\fermion[\NE\REG](\photonbackx,\photonbacky)[3000]
\drawarrow[\NE\ATBASE](\pmidx,\pmidy)
\drawline\fermion[\SE\REG](\photonbackx,\photonbacky)[3000]
\drawarrow[\NW\ATBASE](\pmidx,\pmidy)
\put(21500,12000){$e^-$}
\put(21500,3000){$e^+$}
\put(33000,12000){$Z^0$}
\put(38000,3000){$Z^0$}
\put(39300,10400){$b$}
\put(39300,4900){$\bar b$}
\put(28000,2000){$(18)$}
\end{picture}

\vskip 1.0cm
\centerline{\bf\Large Figure 1 (Continued)}

\vfill

\begin{thebibliography}{1}

\bibitem{unitarity} M.~Veltman, \pl B70 1977 253;\\
B.W.~Lee, C.~Quigg and G.B.~Thacker, {\it Phys. Rev. Lett.}
                    {\bf 38} (1977) 883; {\it Phys. Rev.} {\bf D16} (1977)
1519.

\bibitem{limSM} ALEPH Collaboration, \prep 216 1992 253;\\
                DELPHI Collaboration, \np B373 1992 3;\\
                L3 Collaboration, \pl B303 1993 391; \\
                OPAL Collaboration, \pl B253 1991 511.

\bibitem{guide} J.F.~Gunion, H.E.~Haber, G.L.~Kane and S.~Dawson,
                {\it ``The Higgs Hunter Guide''}, Addison-Wesley, Reading MA,
1990.

\bibitem{LHC} Proceedings of the ``{\it Large Hadron Collider Workshop}'',
Aachen, 4-9 October
              1990, eds. G.~Jarlskog and D.~Rein, Report CERN 90-10, ECFA
90-133, Geneva, 1990.

\bibitem{SSC} Proceedings of the ``{\it Summer Study on High Energy Physics in
the 1990s}'',
              ed. S.~Jensen, Snowmass, Colorado, 1988;\\
              Proceedings of the ``{\it 1990 Summer Study on High Energy
Physics:
              Research Directions for the Decade}'',
              ed. E.L.~Berger, Snowmass, Colorado, 1990.

\bibitem{Tevatron} J.F.~Gunion and T.~Han, \preprint\ UCD-94-10, April 1994.

\bibitem{LepII} Proc. of the ECFA workshop on LEP 200, A. Bohm and W. Hoogland
eds.,
                Aachen FRG, 29 Sept.-1 Oct. 1986, CERN 87-08.

\bibitem{NLC} Proceedings of the Workshop ``{\it Phy\-sics and
Ex\-pe\-ri\-men\-ts with Li\-ne\-ar Col\-li\-ders}'',
              Sa\-ar\-isel\-k\"a, Fin\-land, 9-14 Sep\-tem\-ber 1991, eds.
R.~Orawa, P.~Eerola and M.~Nordberg,
              World Scientific Publishing, Singapore, 1992.

\bibitem{ee500} Proc. of the Workshop ``{\it $e^+e^-$ Collisions at 500 GeV.
The Physics Potential}\ '',
      		Munich, Annecy, Hamburg, 3-4 February 1991, ed. P.M.~Zerwas, DESY pub.
92-123A/B,
      		August 1992.

\bibitem{LC92} Proc. of the ECFA workshop on ``{\it $e^+e^-$ Linear
               Colliders LC92}'', R.~Settles ed., Garmisch Partenkirchen, 25
July-2 Aug.
               1992, MPI-PhE/93-14, ECFA 93-154.

\bibitem{JLC} Proc. of the I Workshop on Japan Linear Collider (JLC), KEK
1989,
              KEK-Report 90-2;\\
              Proc. of the II Workshop on Japan Linear Collider (JLC), KEK
1990,
              KEK-Report 91-10.

\bibitem{gamgam} C.~Seez et al., in ref.\cite{LHC}.

\bibitem{gny} S.L.~Glashow, D.V.~Nanopoulos and A.~Yildiz, {\it Phys. Rev.}
              {\bf D18} (1978) 1724.

\bibitem{wh} R.~Kleiss, Z.~Kunszt and W.J.~Stirling, {\it Phys. Lett.} {\bf
B253} (1991) 269;\\
      	     M.L.~Mangano, SDC Collaboration note SSC-SDC-90-00113.

\bibitem{rwnz} R.~Raitio and W.W.~Wada, {\it Phys. Rev.} {\bf D19} (1979)
941;\\
               J.N.~Ng and P.~Zakarauskas, {\it Phys. Rev.} {\bf D29} (1984)
876.

\bibitem{tth} J.F.~Gunion, {\it Phys. Lett.} {\bf B261} (1991) 510;\\
              W.J.~Marciano and F.E.~Paige, {\it Phys. Rev. Lett.} {\bf 66}
(1991) 2433;\\
      	      A.~Ballestrero and E.~Maina, {\it Phys. Lett.} {\bf B268} (1992)
437;\\
              Z.~Kunzst, Z.~Tr\'ocs\'anyi and W.J.~Stirling, {\it Phys. Lett.}
{\bf B271} (1991) 247;\\
              D.J.~Summers, {\it Phys. Lett.} {\bf B277} (1992) 366.

\bibitem{BCDKZ} V.~Barger, K.~Cheung, A.~Djouadi, B.A.~Kniehl and P.M.~Zerwas,
                \pr D49 1994 79.

\bibitem{bremSM} J.D.~Bjorken, Proceedings of the ``{\it Summer Institute on
Particle
                 Physics}'', {\it SLAC Report} 198 (1976);\\
                 B.W.~Lee, C.~Quigg and H.B.~Thacker, \pr D16 1977 1519;\\
                 J.~Ellis, M.K.~Gaillard and D.V.~Nanopoulos, \np B106 1976
292;\\
                 B.L.~Ioffe and V.A.~Khoze, {\it Sov. J. Part. Nucl.} {\bf 9}
(1978) 50.

\bibitem{fusionSM} D.R.T.~Jones and S.T.~Petkov, \pl B84 1979 440;\\
                   R.N.~Chan and S.~Dawson, \pl B136 1984 196;\\
                   K.~Hikasa, \pl B164 1985 341;\\
                   G.~Altarelli, B.~Mele and F.~Pitolli, \np B287 1987 205;\\
                   B.~Kniehl, {\it preprint} DESY 91-128, 1991.

\bibitem{BCKP} V.~Barger, K.~Cheung, B.A.~Kniehl and R.J.~Phillips,
               {\it Phys. Rev.} {\bf D46} (1992) 3725.

\bibitem{4jet} K.~Hagiwara, J.~Kanzaki and H.~Murayama,
               {\it Durham Univ. Report} No. DTP-91-18, 1991.

\bibitem{SDC} {\it Solenoidal Detector Collaboration Technical Design Report},
              E.L.~Berger {\it et al.}, {\it Report} SDC-92-201, SSCL-SR-1215,
1992.

\bibitem{btagg} J.~Dai, J.F.~Gunion and R.~Vega, \prl 71 1993 2699;\\
                J.~Dai, J.F.~Gunion and R.~Vega, \pl B315 1993 355.

\bibitem{eezh} A. Ballestrero, E. Maina and S. Moretti, \pl B335 1994 460.

\bibitem{tag} H.~Borner and P.~Grosse-Wiesmann, in ref.~\cite{ee500}.

\bibitem{GHS} P.~Grosse-Wiesmann, D.~Haidt and H.J.~Schreiber, in
              ref.~\cite{ee500}.

\bibitem{hz} K.~Hagiwara and D.~Zeppenfeld,
             {\it Nucl. Phys.} {\bf B274} (1986) 1.

\bibitem{BRS} K. Fujikawa, B.W. Lee and A.I. Sanda, {\it Phys. Rev.}
              {\bf D6} (1972) 2923;\\
              C. Becchi, A. Rouet and B. Stora, {\it Comm. Math. Phys.}
              {\bf 42} (1975) 127; {\it Ann. Phys.} {\bf 98} (1976) 287;\\
              B.W. Lee, C. Quigg and H.B. Thacker, {\it Phys. Rev}
              {\bf D16} (1977) 1519;\\
              M.S. Chanowitz and M.K. Gaillard, {\it Nucl. Phys.}
              {\bf B261} (1985) 379;\\
              G.J. Gounaris, R. K\"ogerler and H. Neufeld,
              {\it Phys. Rev.} {\bf D34} (1986) 3257.

\bibitem{tim} T.~Stelzer and W.F.~Long, {\it Comp. Phys. Comm.} {\bf 81}
              (1994) 357.

\bibitem{helas} E.~Murayama, I.~Watanabe and K.~Hagiwara, HELAS: HELicity
                Amplitude Subroutines for Feynman Diagram Evaluations,
                {\it KEK Report} 91-11, January 1992.

\bibitem{vegas} G.P.~Lepage, {\it Jour. Comp. Phys.} {\bf 27} (1978) 192.

\bibitem{ISR} T.~Barklow, P.~Chen and W.~Kozanecki, in ref.~\cite{ee500}.

\end{thebibliography}
\end{document}